\documentclass[preprint,showpacs,preprintnumbers,amsmath,amssymb]{revtex4}

\usepackage{graphicx}% Include figure files
\usepackage{dcolumn}% Align table columns on decimal point
\usepackage{bm}% bold math

\begin{document}

%\preprint{APS/123-QED}

\title{Tension-Induced Morphological Transition in
Mixed Lipid Bilayers}% Force line breaks with \\

\author{S. Komura}
\author{N. Shimokawa}
\affiliation{%
Department of Chemistry, Faculty of Science,
Tokyo Metropolitan University, Tokyo 192-0397, Japan
}%
\author{D. Andelman}
\affiliation{%
School of Physics and Astronomy, Raymond and Beverly Sackler Faculty
of Exact Sciences, Tel Aviv University, Ramat Aviv 69978, Tel Aviv,
Israel
}%

\date{Nov. 20, 2005}% It is always \today, today,
             %  but any date may be explicitly specified

%\begin{abstract}
%\end{abstract}
\pacs{87.16.Dg}

%\keywords{Suggested keywords}%Use show keys class option if keyword
                              %display desired
\maketitle

Raft domains in biological cell membranes are associated with
membrane signaling pathways and have attracted great deal of
interest in recent years~\cite{SI,BL}. Due to the complexity of
biological membranes, a minimal model to investigate rafts (or more
precisely, domain formation) consists of three-component lipid
bilayers containing saturated and unsaturated lipids (e.g.,
sphigomyelin and DOPC) as well as cholesterol. These three-component
systems have been investigated both at the air-water interface and
in artificially vesicles~\cite{VK,BHW}. It was observed that these
multi-component systems exhibit complex phase separation and
appearance of domains.

Recently, Rozovsky et al.\ reported on the morphology and dynamics
of superstructures in such a ternary mixture~\cite{RKG}. A giant
unilamellar vesicle is placed in contact with a lipid bilayer that
is supported on top of a silica surface. 
Fluorescence microscopy is used to monitor the creation and temporal
evolution of domains on the adhering vesicle. In a number of cases,
the upper  part of the vesicles  exhibit a stripe pattern of quite
well-characterized stripe thickness but without any long-range
stripe orientation. It is important to note that due to the large
size of the vesicle, the vesicle upper part is far and in no contact
with the adhering surface. As the adhesion process evolves in time,
the stripe pattern transforms quite abruptly into another type
consisting of an ordered hexagonal lattice of circular domains. The
transformation of the lamellae into circular domains is seen first
as an instability of the stripe tip. Eventually the entire stripes
are replaced by circular domains having roughly the same diameter as
the original stripe width. As time progresses even further, the 
vesicle looks more tense, and the hexagonal pattern of circular 
domains becomes more disordered.

The aim of our Comment is to point out that the observed sequence of
the stripe to hexagonal morphological transition in mixed bilayers
can be attributed to an enhanced membrane surface tension that is
induced by the vesicle adhesion on the solid surface.

It is known that surface tension can induce adhesion of a vesicle
onto a substrate or to another vesicle~\cite{Seifert}. The reason
being that surface tension of the membrane suppresses thermal
fluctuations and promotes its adhesive properties. A self-consistent
treatment of such a problem predicted the conditions for
tension-induced adhesion~\cite{Seifert}. On the other hand, it was
reported that the spreading of red blood cells on a substrate
produces finite surface tension which can even cause them to
rupture~\cite{HLKD}. More recent experiments on the
pulling of nanotubes from adhered vesicles revealed that the
membrane tension strongly increases during elongation~\cite{CCBN}.
In view of these works, we explore the relation between an increase
in surface tension and lateral morphological transitions that can
occur on the membrane plane.

The main ingredient of our model is to introduce a coupling between
local composition on the membrane plane and the local
curvature~\cite{LA}. We do not consider a closed form vesicle of a
finite area and volume. We rather model a planar and extended
membrane. Its shape is described by a displacement variable
$\ell(x,y)$, relative to the reference $x$-$y$ plane. The local
composition (assuming a simple binary mixture of two components) is
similarly denoted by $\phi(x,y)$. The total free energy is written
in terms of these two local fields. For simplicity, we consider only
small undulations above the $x$-$y$  plane in the form of the
following expansion~\cite{LA}:
\begin{equation}
H\{ \ell, \phi \} = \int {\rm d}^2 x \left[ \frac{1}{2} \sigma
(\nabla \ell)^2 + \frac{1}{2} \kappa (\nabla^2 \ell)^2 + \frac{1}{2}
b (\nabla \phi)^2 + \frac{1}{2} a_2 \phi^2 + \frac{1}{4} a_4 \phi^4
+ \Lambda (\nabla^2 \ell) \phi \right], \label{hamiltonian}
\end{equation}
where $\sigma>0$ and $\kappa>0$ are the surface tension and bending
rigidity of the membrane, respectively. The other coefficients, $b$,
$a_2$, $a_4$ and $\Lambda$ are phenomenological ones.  The parameter
$b>0$ is related to the line tension between different domains,
$a_2\sim T-T_{\rm c}$ is the reduced temperature ($T_{\rm c}$ is the
critical temperature). Below the phase transition it is negative and
we need to include in the expansion a positive 4$^{th}$ order term,
$a_4>0$, for stability purposes. Finally, $\Lambda$ is the coupling
constant. This coupling term represents the situation where the
spontaneous curvature of the membrane depends on the local
composition. For simplicity, the other coupling term $(\nabla^4
\ell) \phi$ (higher order in Fourier space) as well as higher-order
coupling terms are omitted here.

Starting from the above free energy, we first integrate out the
$\ell$-field in Fourier space, and the effective free energy is
approximated as the Brazovskii free energy~\cite{LA}:
\begin{equation}
H_{\rm eff} \{ \phi \} \approx \int {\rm d}^2 x \left[
\frac{1}{2} B (\nabla \phi)^2
+ \frac{1}{2} C (\nabla^2 \phi)^2
+ \frac{1}{2} a_2 \phi^2
+ \frac{1}{4} a_4 \phi^4
\right],
\label{effhamiltonian}
\end{equation}
where
\begin{equation}
B \equiv b - \frac{\Lambda^2}{\sigma},~~~~~
C \equiv \frac{\Lambda^2 \kappa}{\sigma^2}.
\label{defBandC}
\end{equation}
In deriving the above expression, we performed an expansion up to
4$^{th}$ order in the wavevector $q$ (or equivalently up to 4$^{th}$
order in differential operator in real space). The homogeneous
disordered phase becomes unstable with respect to the modulated
phase when the coupling constant is large enough so that $\Lambda^2
> b \sigma$. Note that the sign of $\Lambda$ can also be negative
as the results depend only on $\Lambda^2$. The condition to have
finite unstable $q$ vectors is simply that $B<0$.  The most unstable
wavevector is then given by $q^{\ast} = (-B/2C)^{1/2}$.

We recall that the  mean-field free energies of the stripe and the
hexagonal phases have been used in Ref.~\cite{LA}  in order to
construct the membrane phase diagram (see Fig.~1 of Ref.~\cite{LA}).
Only the most unstable wavevector $q^{\ast}$ (single-mode
approximation) is used in the calculation, which is justified for
the region close to the homogeneous critical point. There are four
distinct phases: disordered (D), stripe (S), hexagonal (H), and
inverted hexagonal phases. These phases are separated by two-phase
coexistence regions (first-order phase transitions). The calculated
phase diagram is in agreement with experiments performed  on
mixtures of lipids in Langmuir monolayers at the air-water
interface~\cite{KM}.

In order to see clearly the role of surface tension $\sigma$, we
cannot use directly the phase diagram of Ref.~\cite{LA}. Instead, we
construct  a new phase diagram in terms of two different reduced
variables defined by \cite{commentLA}
\begin{equation}
\Omega \equiv \frac{a_2}{a_4 \phi_0^2},~~~~~ \Sigma \equiv \frac{b
\sigma}{\Lambda^2}. \label{newparameter}
\end{equation}
In the above, $\phi_0=\langle \phi \rangle$ is the average
composition that is fixed by the chemical potential. We note that
$\Omega$ is independent of the surface tension parameter $\sigma$,
and that the only variation of $\sigma$ enters in the second
variable, $\Sigma\sim\sigma$.

\begin{figure}
\includegraphics[scale=0.7]{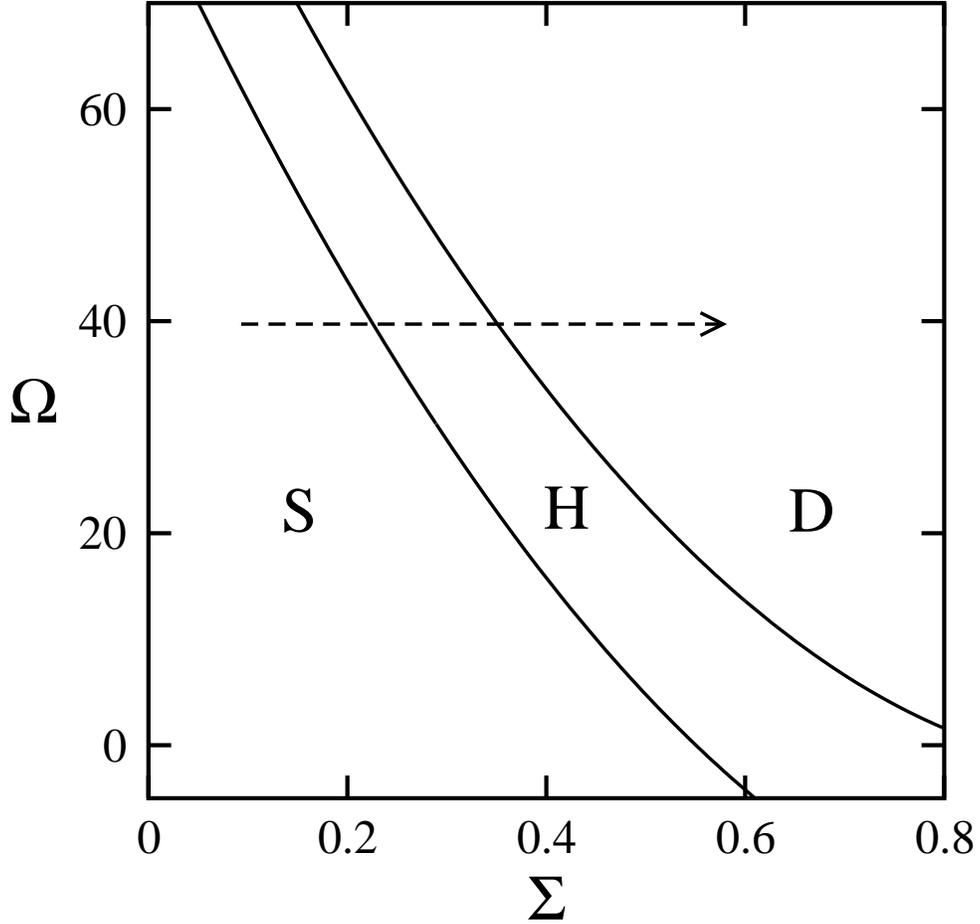}
\caption{Mean-field phase diagram in the ($\Sigma$, $\Omega$) plane
for $M=0.1$. $\Sigma$ represents the reduced surface tension and
$\Omega$ the curvature-composition coupling parameter, as defined in
Eq.~(\ref{newparameter}). There are three different phases: the
stripe phase (S), the hexagonal phase (H), and the disordered
(homogeneous) phase (D). These phases are separated by the
first-order transition lines. For simplicity we show here by solid
lines only the crossover in the free energies while avoiding
plotting two-phase regions. The effect of increasing the surface
tension is represented by the dashed arrow. } \label{fig1}
\end{figure}

Figure~\ref{fig1} is the corresponding phase diagram in the
$(\Omega, \Sigma)$ plane when $M \equiv (4 \kappa a_4)^{1/2} \phi_0
/ \Lambda $ is kept constant. In the figure, we chose $M=0.1$. Regions 
of coexistence are omitted in this phase diagram for clarity purposes.
Thus, the transition lines indicate the locus of points at which the
free energies of different phases cross each other. The stripe phase
(S) appears when $\Sigma\sim\sigma$ is small.  The dashed arrow in
Fig.~\ref{fig1} indicates a horizontal path along which $\Sigma$ (or
equivalently the surface tension $\sigma$) increases for a fixed
value of $\Omega$. Clearly, along this path the stripe phase (S)
transforms into the hexagonal phase (H). As $\Sigma$ increases even
further, there will be another phase transition into the homogeneous
disordered phase (D). Such a sequence of the pattern transformation,
S $\to$ H, upon increasing the surface tension is consistent with
that observed in the mixed lipid bilayers (see Fig.~1 in
Ref.~\cite{RKG}). This is our main result. We consider that the
enhanced surface tension during the adhesion process is responsible
for the morphological evolution of the domain shapes in the adhering
vesicle. It should be noted that although the calculated phase
diagram depends on the value of $M$, the sequence of the modulated
structures remains unchanged.

We discuss now our results and their relationship with experiment.
In  Ref.~\cite{RKG}, the vesicles are shown to become more tensed as
the adhesion process evolves. This results in a reduced area of the
vesicle (see Fig.~1 of Ref.~\cite{RKG}). In the case of mixed
vesicles, the spontaneous curvature arises from the asymmetry in the
composition between the inner and outer leaflets comprising the
bilayer membrane. Such an effect is represented in the last coupling
term of Eq.~(\ref{hamiltonian}). Although the evidence of the
asymmetric composition between the inner and outer leaflets has not
been reported in Ref.~\cite{RKG}, it certainly exists in biological
cells due to an active mechanism.  Asymmetry in the composition
might also occur in situations where the two leaflets experience a
different environment. In an adhering vesicle, for example, only the
outer leaflet is in contact with the surface. More detailed theories
dealing with the bilayer nature of the membranes are given in
Refs.~\cite{KK,KGL}. It should be noted, however, that these works
as well as the present one consider only an averagely flat and
extensive membrane, rather than closed vesicles of finite area and
volume. In the previous experiments of Ref.~\cite{BHW}, similar
stripe and hexagonal patterns have been shown but no similar
temporal evolution was reported. This may be because the vesicles in
Ref.~\cite{BHW} were suspended in an aqueous solution (rather than
adhering on a surface), and their surface tension is not increasing
as function of time. It is also entirely not clear how much the
patterns and domains seen on the vesicular surface are truly
equilibrium ones or just long-lived metastable ones.

Finally, we comment on the dynamical aspect of the observed
morphological transition. In the last decade, kinetics of phase
transitions between different ordered mesophases have been studied
intensively, especially, for block copolymer systems~\cite{QW}.
Among various works, Nonomura and Ohta investigated the
morphological transition between the lamellar and the hexagonal
phases in two dimensions~\cite{NO}. Their dynamical equation is
based on the time-dependent Ginzburg-Landau equation using a free
energy which includes a long-range repulsive interaction term. This
free energy is known to be similar to the one used in the present
work.

In the simulation, the transition is induced by changing the
temperature. Note that in our Fig.~\ref{fig1}, a change in
temperature is simply a vertical scan where $\Omega\sim a_2\sim
T-T_c$ changes, while $\Sigma\sim\sigma$ (as well as $M$ or
$\phi_0$) are kept fixed. The vertical scan can also lead to a
crossover between the stripe and hexagonal phases. In the
experiments~\cite{RKG}, on the other hand, the temperature is kept
fixed and we believe that the origin of the transition relies on the
temporal increase in the surface tension as detailed above. Despite
the difference in the control parameters (either temperature or
surface tension), the time evolution of the structure in the
simulation is very similar to that observed for the lipid mixtures.
For example, the stripes undergo a pearling phenomenon and the
hexagonal domains appear at the grain boundaries of the lamellar
structure~\cite{NO}. We hope that more quantitative measurements of
the surface tension during the adhesion process will shed light on
this intriguing phenomena and will lead to further theoretical
investigations.

%%%%%%%%%%%%%%%%%%%%%%
%  Acknowledgments   %
%%%%%%%%%%%%%%%%%%%%%%

\begin{acknowledgments}
We thank J.\ Groves, Y.\ Kaizuka, P.\ Nassoy, S.\ Rozovsky and 
T.\ Ohta for useful discussions.
This research was partially supported by the Ministry of Education,
Culture, Sports, Science and Technology, Japan under grant
No.\ 15540395, the Israel Science Foundation (ISF) under grant
No.\ 160/05 and the US-Israel Binational Foundation (BSF) under grant
No.\ 287/02.
\end{acknowledgments}

%\bibliography{lipid}% Produces the bibliography via BibTeX.

\begin{references}


\bibitem{SI}
K. Simons and E. Ikonen,
Nature {\bf 387}, 569 (1997).

\bibitem{BL}
D. A. Brown and E. London,
J.\ Bio.\ Chem.\ {\bf 275}, 17221 (2000).

\bibitem{VK}
S. L. Veatch and S. L. Keller,
Phys.\ Rev.\ Lett.\ {\bf 89}, 268101 (2002).

\bibitem{BHW}
T. Baumgart, S. T. Hess, and W. W. Webb,
Nature {\bf 425}, 821 (2003).

\bibitem{RKG}
S. Rozovsky, Y. Kaizuka, and J. T. Groves,
J.\ Am.\ Chem.\ Soc.\ {\bf 127}, 36 (2005).

\bibitem{Seifert}
U. Seifert, Phys.\ Rev.\ Lett.\ {\bf 74}, 5060 (1995).

\bibitem{HLKD}
A. Hategan, R. Law, S. Kahn, D. E. Discher,
Biophys.\ J.\ {\bf 85}, 2746 (2003).

\bibitem{CCBN}
D. Cuvelier, N. Chiaruttini, P. Bassereau and P. Nassoy,
Europhys.\ Lett.\ {\bf 71}, 1015 (2005).

\bibitem{LA}
S. Leibler and D. Andelman,
J.\ Phys.\ (Paris) {\bf 48}, 2013 (1987).

\bibitem{commentLA}
In Ref.~\cite{LA}, the phase diagram was presented in terms of the
reduced temperature $\varepsilon \equiv {4 C a_2}/{B^2}$ and the
reduced average concentration $m \equiv \left({4 C a_4}/{B^2}
\right)^{1/2} \phi_0$, where the numerical prefactors of ``2'' in
Eq.~(10) of Ref.~\cite{LA} are corrected. In terms of the above
parameters, we have $\Omega=\varepsilon/m^2$ and $\Sigma= 1 -
(M/m)$, where $M$ is defined in the text.

\bibitem{KM}
S. L. Keller and H. M. McConnell,
Phys.\ Rev.\ Lett.\ {\bf 82}, 1602 (1999).

\bibitem{KK}
H. Kodama and S. Komura,
J.\ Phys.\ II {\bf 3}, 1305 (1993).

\bibitem{KGL}
P. B. S. Kumar, G. Gompper, and R. Lipowsky,
Phys.\ Rev.\ E {\bf 60}, 4610 (1999).

\bibitem{QW}
S. Qi and Z.-G. Wang,
Phys.\ Rev.\ Lett.\ {\bf 76}, 1679 (1996);
Phys.\ Rev.\ E {\bf 55}, 1682 (1997).

\bibitem{NO}
M. Nonomura and T. Ohta,
J.\ Phys.\ Soc.\ Jpn.\ {\bf 70}, 927 (2001).

\end{references}
%%%%%%%%%%%%%%%%%%%%%%%%%%%%%%%%%%%%%%%%%%%%%%%%%%%%%%%%%%%%%%%%%%%%%
%  References  %%%%%%%%%%%%%%%%%%%%%%%%%%%%%%%%%%%%%%%%%%%%%%%%%%%%%%
%%%%%%%%%%%%%%%%%%%%%%%%%%%%%%%%%%%%%%%%%%%%%%%%%%%%%%%%%%%%%%%%%%%%%

%%%%%%%%%%%%%%%%%%%%%%%%%%%%%%%%%%%%%%%%%%%%%%%%%%%%%%%%%%%%%%%%%%%%%
%  Tables & Figures  %%%%%%%%%%%%%%%%%%%%%%%%%%%%%%%%%%%%%%%%%%%%%%%%
%%%%%%%%%%%%%%%%%%%%%%%%%%%%%%%%%%%%%%%%%%%%%%%%%%%%%%%%%%%%%%%%%%%%%

\clearpage

\end{document}